\documentclass{phc-proc4-auth}
\usepackage{amssymb}
\usepackage[dvips]{graphicx}
\newcommand{\ped}[1]{\ensuremath{_{\rm #1}}}

\begin{document}
\begin{frontmatter}

\title{The determination of the electron-phonon interaction from
       tunneling data in the two-band superconductor MgB$_2$}

\author[To]{D.Daghero\corauthref{cor}}
\corauth[cor]{Corresponding author}
\ead{dario.daghero@infm.polito.it},
\author[To]{R.S. Gonnelli},
\author[To]{G.A. Ummarino},
\author[2]{O.V. Dolgov},
\author[2]{J. Kortus},
\author[3]{A.A. Golubov},
\author[4]{S.V. Shulga}

\address[To]{INFM - Dipartimento di Fisica, Politecnico di Torino,
Corso Duca degli Abruzzi 24, 10129 Torino, Italy}

\address[2]{Max-Planck Institut f\"{u}r
        Festk\"{o}rperforschung, Stuttgart, Germany}

\address[3]{Department of
        Applied Physics, University of Twente, 7500 AE Enschede, The
        Netherlands}
\address[4]{Institute f\"{u}r Festk\"{o}rper- und
        Werkstofforschung, Dresden, Germany}

\begin{abstract}
We calculate the tunneling density of states (DOS) of MgB$_2$ for
different tunneling directions, by directly solving the real-axis,
two-band Eliashberg equations (EE). Then we show that the numeric
inversion of the standard \emph{single-band} EE, if applied to the
DOS of the \emph{two-band} superconductor MgB$_2$, may lead to
wrong estimates of the strength of certain phonon branches (e.g.
the $E_{2g}$) in the extracted electron-phonon spectral function
$\alpha^{2}F(\omega)$. The fine structures produced by the
two-band interaction turn out to be clearly observable only for
tunneling along the $ab$ planes in high-quality single crystals.
The results are compared to recent experimental data.
\end{abstract}
\end{frontmatter}

There is a growing consensus that superconductivity in MgB$_2$ is
driven by the electron-phonon interaction (EPI) \cite{MA}. The
idea of multiband superconductivity in MgB$_{2}$ \cite%
{liu,br,Choi,mazinimp} is supported by many recent
experimental results from tunneling \cite{IavaronePRL02,Esk}, point contact %
\cite{SzaboPRL01,Gonnelli} and specific heat measurements
\cite{BouquetPRL01}. These data give direct evidence that the
superconducting gap has two different values: $\Delta _{\sigma }$
for the two quasi-two-dimensional $\sigma$ bands, and $\Delta
_{\pi }$ for the 3D $\pi$ bands \cite{liu,br}. According to most
calculations \cite{liu,Choi,kong}, the EPI or, equivalently, the
Eliashberg spectral function $\alpha ^{2}F(\omega )$ (EF) is
dominated by the optical boron bond-stretching $E_{2g}$ phonon
branch around 60 - 70 meV.

At present, the main experimental tool for the determination of the EPI
in superconductors is tunnel spectroscopy. Unfortunately, the
standard single-band procedure to obtain the EF from the first
derivative of the tunneling current is restricted to a
momentum-independent $s$-wave order parameter and cannot be used in
anisotropic superconductors like MgB$_2$. Nevertheless, there has
been a recent attempt to obtain the EPI in MgB$_{2}$ by using this
standard approach \cite{d1a}.

In the present paper we clarify what information can be obtained by
using this \emph{single-band} standard procedure if one applies it
to a \emph{two-band} superconductor.

The parameters for the two-band model used in this work are based
on first-principle electronic structure calculations \cite{kong}.
The four interband and intraband electron-phonon spectral
functions $\alpha ^{2}F_{ij}(\omega )$ (where $i,j=\pi ,\sigma $)
and the Coulomb pseudopotential matrix $\mu _{ij}^{\ast }$ are the
basic input for the two-band Eliashberg theory. The $\sigma\sigma
$ EPI is dominated by the optical boron bond-stretching $E_{2g}$
phonon mode. For the other channels there are also important
contributions from low-frequency (30-40 meV) and from
high-frequency phonon modes ($\simeq$ 90 meV).

\begin{figure}[t]
\begin{center}
\includegraphics[width=1.0\columnwidth]{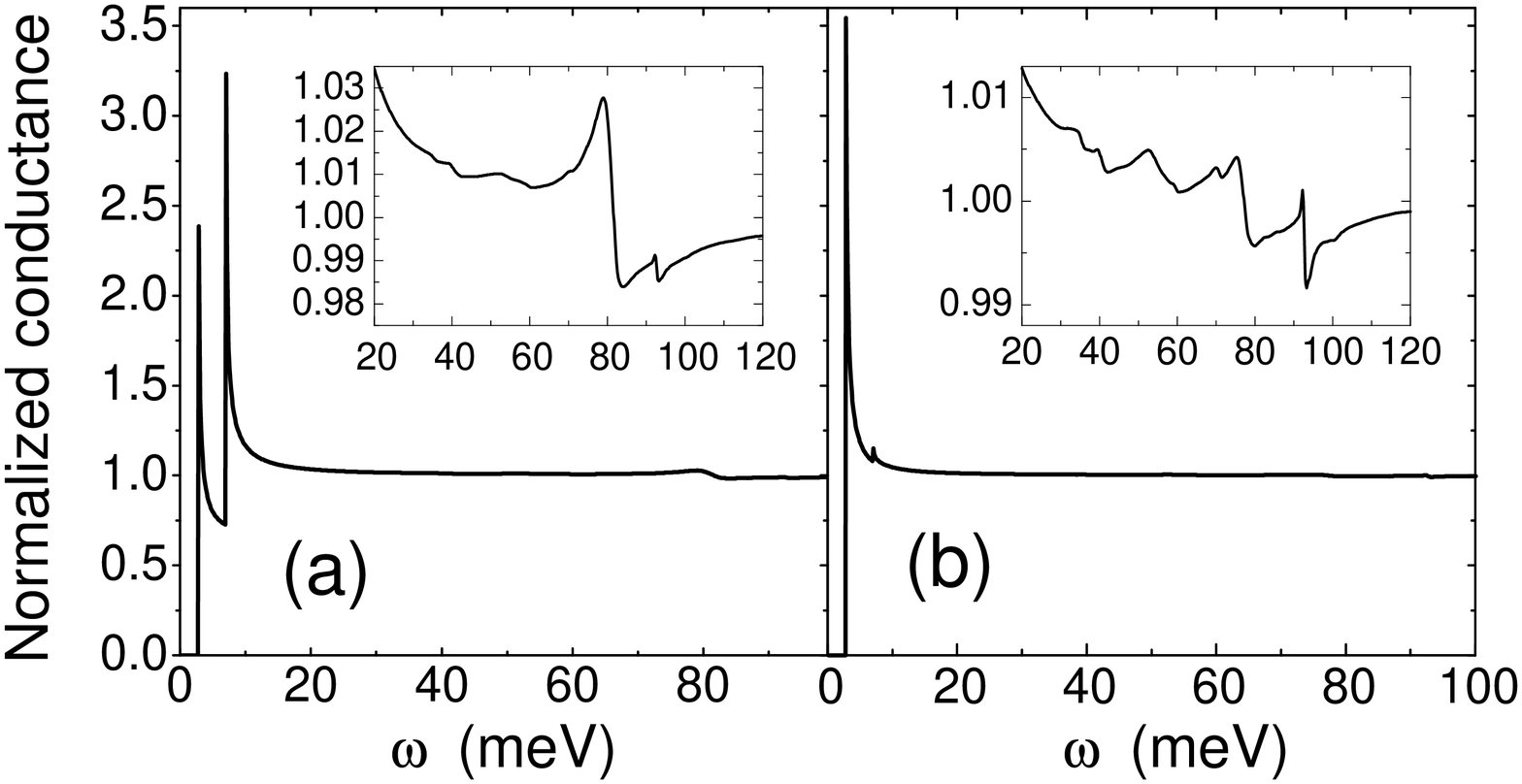}
\end{center}
\caption{The calculated tunneling DOS at T=0 K in the
\textit{ab}-plane (a) and along the \textit{c}-axis (b). The two
insets show the fine structures of the tunneling DOS due to the
electron-phonon interaction. } \label{fig:1}
\end{figure}

We obtained the theoretical conductance curves of MgB$_{2}$ for
different tunneling directions by solving the corresponding
two-band Eliashberg equations (EE) \cite{br,dolgov} in the
real-axis formulation. The only free parameter is the
normalization constant $\mu $ in the Coulomb pseudopotential
matrix which is fixed in order to reproduce the experimental
$T_{c}=39.4$ K. Figures \ref{fig:1} (a) and (b) show the
calculated conductances for tunneling along the \textit{ab}-plane
and the \textit{c}-axis directions, respectively. According to
Ref.\cite{br}, these conductances are a weighed sum of the
$\sigma$- and $\pi$-band contributions to the DOS, whose weights
$w\ped{\sigma}$ and $w\ped{\pi}$ are determined by the
corresponding Fermi velocities in the bands and by the angle of
the tunneling current with respect to the \textit{ab} plane. The
values of these weights are $w\ped{\sigma}$=0.33 and
$w\ped{\pi}$=0.67 for tunneling along the \textit{ab} plane,
$w\ped{\sigma}$=0.01 and and $w\ped{\pi}$=0.99 for tunneling along
the \textit{c} axis. It is clear that the contribution of the $\pi
$ band is always dominant even if tunneling is almost in the
\textit{ab} plane. The fine structures due to the EPI have maximum
amplitudes of the order of 0.5\% and 2-3\% for measurements along
the \textit{c}-axis and in the \textit{ab}-plane, respectively
(see the two insets of Fig. \ref{fig:1}).

In order to test the reliability of the numeric inversion of the
standard \emph{single-band} Eliashberg equations when applied to a
\emph{two-band} superconductor, we used the calculated tunneling
DOS shown in Fig. \ref{fig:1} as an input. Of course, in this
case, the inverted spectral function should correspond to a
mixture of the $\sigma$- and $\pi$-band contributions. Figures 2
(a) and (b) show the results of the inversion of the calculated
tunneling DOS of Fig.1 (a) and (b), respectively,  using a
standard single-band code (solid lines). The effect of the
single-band EE inversion is estimated by a least-square fit of the
inverted spectral functions with the weights for $\alpha
^{2}F_{\sigma }=\alpha ^{2}F_{\sigma \sigma }+\alpha ^{2}F_{\sigma
\pi }$ and $\alpha ^{2}F_{\pi }=\alpha ^{2}F_{\pi \pi }+\alpha
^{2}F_{\pi \sigma }$ as free parameters. These fits are shown by
the dashed curves in Fig.2 and correspond to $\alpha
^{2}F_{ab}^{fit} \simeq 0.31\alpha ^{2}F_{\sigma } +0.16\alpha
^{2}F_{\pi } $ and $\alpha ^{2}F_{c}^{fit} \simeq 0.01\alpha
^{2}F_{\sigma } +0.99\alpha ^{2}F_{\pi } $.

Tunneling along the $c$-direction practically corresponds to a
single-band case and, as expected, the inversion properly
reproduces the $\alpha ^{2}F_{\pi}$. In contrast, in the case of
tunneling along the $ab$-direction, the $\sigma$-band spectral
functions play an essential (and amplified) role since about 33\%
of the $\sigma$-band contribution in the tunneling DOS corresponds
to a contribution of about 66\% of $\alpha ^{2}F_{\sigma }$ in the
effective $\alpha ^{2}F_{ab}$.

The above results show that, in polycrystalline samples and in
non-oriented films (where the $\pi $ band dominates the tunneling
current), the inversion of the experimental DOS gives mainly
information about $\alpha ^{2}F_{\pi }(\omega )$, and thus leads
to underestimate the strength of the $E_{2g}$ phonon mode that is
essential for superconductivity. This might explain the results
reported in Ref.\cite{d1a}, where the dominant role of the
$E_{2g}$ phonon mode was questioned. Thus, only separate studies
of very low-temperature \textit{ab}-plane and \textit{c}-axis
tunneling conductances in high-quality single crystals in clean
limit might allow a quantitative estimate of the EPI. These
studies should provide a crucial test for the results of the
two-band model, even though the fine structures in the DOS due to
the EPI are very likely to be experimentally observable only for
tunneling current parallel to the $ab$ planes.

\begin{figure}[t]
\begin{center}
\includegraphics[width=1.0\columnwidth]{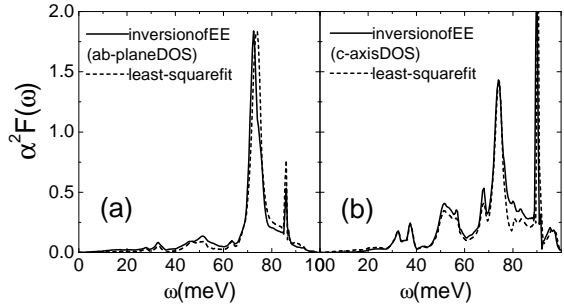}
\end{center}
\caption{The Eliashberg functions for tunneling along the
\textit{ab}-plane (a) and along the \textit{c-}axis (b) obtained
from the curves of Fig.1 by inversion of the single-band
Eliashberg equations (solid line). The dashed lines represent the
least-square fits with two free parameters as explained in the
text. } \label{fig:2}
\end{figure}

\end{document}